\documentclass[useAMS,usenatbib]{mnras}
\usepackage{graphicx,graphics}
\usepackage{amsmath}

\newcommand{\diff}{\mathrm{d}}
\newcommand{\tgw}{t_{\mathrm{gw}}}
\newcommand{\ncand}{N_{\mathrm{cand}}}
\newcommand{\ncandsin}{N_{\mathrm{cand,noex}}}
\newcommand{\ncandex}{N_{\mathrm{cand,ex}}}
\newcommand{\nin}{N_{\mathrm{in}}}
\newcommand{\nout}{N_{\mathrm{out}}}
\newcommand{\rc}{R_{\mathrm{c}}}
\newcommand{\rhnobh}{R_{\mathrm{h,noBH}}}
\newcommand{\rh}{R_{\mathrm{h}}}
\newcommand{\rhbh}{R_{\mathrm{h,BH}}}
\newcommand{\mbbh}{m_{\mathrm{BBH}}}
\newcommand{\mbhc}{\langle m_{\mathrm{BH,c}} \rangle}

\newcommand{\Mbh}{M_\mathrm{BH}}
\newcommand{\nbhc}{N_{\mathrm{BH,c}}}
\newcommand{\rlagrcbh}{R_{\mathrm{lagr,BH,10\%}}}
\newcommand{\rhobh}{\rho_{\mathrm{BH}}}
\newcommand{\sigmabh}{\sigma_{\mathrm{BH}}}
\newcommand{\vesc}{v_{\mathrm{esc}}}
\newcommand{\tms}{t_{\mathrm{ms}}}
\newcommand{\trh}{t_{\mathrm{rh}}}
\newcommand{\trha}{t_{\mathrm{rh,1}}}
\newcommand{\trhb}{t_{\mathrm{rh,BH}}}
\newcommand{\Eb}{E_{\mathrm{BH}}}
\newcommand{\tmerge}{t_{\mathrm{merge}}}
\newcommand{\nk}{n_{\mathrm{k}}}
\newcommand{\mk}{m_{\mathrm{k}}}
\newcommand{\vk}{v_{\mathrm{k}}}
\newcommand{\vave}{\langle v \rangle}
\newcommand{\n}{\langle n \rangle}
\newcommand{\m}{\langle m \rangle}

\title[Impact of IMFs on GW sources]{Impact of initial mass functions on the dynamical channel of gravitational wave sources}
\author[Long Wang et al.]{Long Wang, $^{1,2}$\thanks{E-mail:long.wang@astron.s.u-tokyo.ac.jp}
  Michiko S. Fujii,$^{1}$
  Ataru Tanikawa$^{3}$
  \\
$^{1}$Department of Astronomy, School of Science, The University of Tokyo, 7-3-1 Hongo, Bunkyo-ku, Tokyo, 113-0033, Japan \\
$^{2}$RIKEN Center for Computational Science, 7-1-26 Minatojima-minami-machi, Chuo-ku, Kobe, Hyogo 650-0047, Japan\\
$^{3}$Department of Earth Science and Astronomy, The University of Tokyo, Japan \\
}
\begin{document}

\date{Accepted --. Received --; in original form --}

\pagerange{\pageref{firstpage}--\pageref{lastpage}} \pubyear{2019}

\maketitle

\label{firstpage}

\begin{abstract}
Dynamically formed black hole (BH) binaries (BBHs) are important sources of gravitational waves (GWs). 
Globular clusters (GCs) provide a major environment to produce such BBHs, but the total mass of the known GCs is small compared to that in the Galaxy; thus, the fraction of BBHs formed in GCs is also small. 
However, this assumes that GCs contain a canonical initial mass function (IMF) similar to that of field stars. 
This might not be true because several studies suggest that extreme dense and metal-poor environment can result in top-heavy IMFs, where GCs may originate. 
Although GCs with top-heavy IMFs were easily disrupted or have become dark clusters, the contribution to the GW sources can be significant. 
Using a high-performance and accurate $N$-body code, \textsc{petar}, we investigate the effect of varying IMFs by carrying out four star-by-star simulations of dense GCs with the initial mass of $5\times10^5 M_\odot$ and the half-mass radius of $2$~pc. 
We find that the BBH merger rate does not monotonically correlate with the slope of IMFs. Due to a rapid expansion, top-heavy IMFs lead to less efficient formation of merging BBHs. The formation rate continuously decreases as the cluster expands because of the dynamical heating caused by BHs. 
However, in star clusters with a top-heavier IMF, the total number of BHs is larger, and therefore, the final contribution to merging BBHs can still be more than from clusters with the standard IMF, if the initial cluster mass and density is higher than those used in our model. 

\end{abstract}


\section{Introduction}



After LIGO/VIRGO detected gravitational wave (GW) events from mergers
of stellar-mass black holes (BHs) and neutron stars (NSs)
\citep{Abbott2019,Abbott2020}, many studies have investigated the
origins of these events: isolated binaries through common envelope
evolution \citep[e.g.][]{Giacobbo2018,Belczynski2020}, through
chemically homogeneous evolution
\citep[e.g.][]{Marchant2016,Mandel2016} and stable mass transfer
\citep[e.g.][]{Kinugawa2014,Tanikawa2020}, hierarchical stellar
systems \citep[e.g.][]{Antonini2014}, open clusters
\citep[e.g.][]{Ziosi2014,Kumamoto2019,DiCarlo2019,Banerjee2020a}, and
galactic centers \citep[e.g.][]{OLeary2009}. 
Dense stellar systems like globular clusters (GCs) are considered to provide conducive environments to form GW progenitors via few-body dynamical interactions
\citep{PortegiesZwart2000,Downing2010,Tanikawa2013,Bae2014,Rodriguez2016a,Rodriguez2016b,Fujii2017,Askar2017,Park2017,Samsing2018,Hong2020}.
Although this dynamical channel can efficiently produce GW events, the total contribution seems to be less than the events driven by the binary stellar evolution, because the total mass of GCs is a small fraction of the total galactic stellar mass.  
However, it is assumed that the initial mass function (IMF) of the GCs is the same as that of the field stars, but it might not be true because we do not have sufficient observational constraints on the IMF of the GCs yet.

Recently, several observational evidences indicate that extreme dense star forming regions may have top-heavy IMFs, such as the Arches cluster in the Galactic center and 30~Doradus in the Large Magellanic Cloud, where the heavy-end of IMFs have exponent, $\alpha \approx -1.7 \sim -1.9$, \citep{Lu2013,Schneider2018,Hosek2019}.
Therefore, it is natural to expect that the old and massive GCs in the Milky Way may also contain top-heavy IMFs since its birth environment is very different from the present-day one.
\cite{Zonoozi2016} and \cite{Haghi2017} found the indirect evidence that a top-heavy IMF can explain the observed trend of metallicity and mass-to-light ratio among the GCs in M31.
On the galactic scale, \cite{Zhang2018} found that for high-redshift ($z\sim 2-3$) star-burst galaxies, a top-heavy (integrated) IMF is necessary to explain their star formation rate.

Meanwhile, there is a puzzle related to the phenomenon of multiple stellar populations in GCs, called the ``mass budget problem''.
Observations show that several GCs have more than half of enriched stellar populations \citep{Milone2017}.
The stellar evolution model to explain the formation of element-enriched stars, especially the AGB scenario, cannot produce sufficient materials to form such a large fraction of young populations \citep[][and references there in]{Bastian2018}.
A top-heavy IMF is a natural solution for this problem \citep[e.g.][]{Wang2020a}.

With top-heavy IMFs, the strong wind mass loss from massive stars in the first 100 Myr significantly affects the density of the systems.
Subsequently, the massive stars leave a large number of BHs in the star clusters, and they will also have a strong impact on the long-term evolution of the clusters.
Indeed, it has been shown that GCs with top-heavy IMFs are much easier to expand and be disrupted \citep{Chatterjee2017,Giersz2019,Wang2020b,Weatherford2021}.

Therefore, the dynamical evolution of GCs provides a strong constraint on the shape of IMFs.
By using semi-analytic models, \cite{Marks2012} suggested that the shape of IMFs in GCs might depend on the metallicity and initial cloud density. Their model, however, ignored the dynamical impact of BHs, and thus, they overestimated the slopes of IMFs for observed GCs.
By comparing with scaled $N$-body models, \cite{Baumgardt2017} also argued that 35 observed dense GCs might not have top-heavy IMFs.
However, GCs with top-heavy IMFs might have existed in the past but have already disappeared or have become dark clusters \citep{Banerjee2011}.
Thus, they cannot be directly observed today.
Considering that a large number of BHs existed there, the contribution of binary black hole (BBHs) mergers might be significant.

\cite{Chatterjee2017}, \cite{Giersz2019} and \cite{Weatherford2021} have performed Monte-Carlo simulations of GCs to study the effect of top-heavy IMFs on the survival of GCs and the BBH mergers. 
However, the Monte-Carlo method has not been fully tested for the condition of top-heavy IMFs, wherein a large fraction of BHs exist.
\citet{Rodriguez2016} compared the Monte-Carlo \citep[the \textsc{cmc} code; e.g. ][]{Joshi2000} and the direct $N$-body methods \citep[the \textsc{nbody6++gpu} code;][]{Wang2015} for million-body simulations of (Dragon) GCs \citep{Wang2016}. In this comparison, the \textsc{cmc} simulation shows a short-period oscillation of the BH core radius. This does not appears in the direct $N$-body model.
Such a short-period oscillation is connected to the formation of BH binaries and the interaction between the binaries with other BHs in the core. 
Since the Dragon models have a low-density and only pass $<2$ initial half-mass relaxation time,
it is unclear that for mode dense GCs, whether such a different behaviour of core evolution can result in different formation and evolution of BH binaries. 
Besides, \cite{Rodriguez2018} showed that when the BH number is large, the \textsc{cmc} simulations show a different core radius compared to that of the direct $N$-body simulation. 
It is unclear as to how such a large difference could exist in the case of top-heavy IMFs. 

There are also studies using direct $N$-body methods, which have no approximation on the gravitational interaction, unlike the case of the Monte-Carlo method.
\citet{Wang2020b} studied the effect of top-heavy IMFs on the survival of star clusters, but used a simplified two-component models without stellar evolution.
\cite{Haghi2020} carried out a group of $N$-body models, but without a realistic number of stars like in the GCs. 
These models properly integrated individual stellar orbits and obtained correct dynamical evolution of GCs, but they cannot be used to study the BBH mergers.
Therefore, it is necessary to carry out accurate star-by-star $N$-body simulations to study the GCs with top-heavy IMFs.


In this work, we carry out four $N$-body simulations of GCs with the sufficient mass ($5\times 10^5 M_\odot$) and density (the half-mass radius is 2~pc). IMFs with different slopes are used.
In Section~\ref{sec:petar}, we describe the $N$-body tool, \textsc{petar}, used in this study.
The initial conditions of models are presented in Section~\ref{sec:init}.
We describe the BBHs mergers and the dynamical evolution of the models in Section~\ref{sec:result}.
Finally, we discuss our results and draw conclusions in Sections~\ref{sec:discussion} and \ref{sec:summary}.

\section{Methods}
\label{sec:method}

\subsection{\textsc{Petar} code\label{sec:petar}}

We use the \textsc{petar} code \citep{Wang2020d} to develop the $N$-body models of GCs.
This is a hybrid code that combines the particle-tree particle-particle \citep{Oshino2011} and slowdown algorithm regularization \citep{Wang2020c} methods.
Such a combination reduces the computing cost compared to the direct $N$-body method while maintaining sufficient accuracy to deal with close encounters and few-body interactions.
The code is based on the \textsc{fdps} framework, which can achieve a high performance with the multi-process parallel computing \citep{Iwasawa2016,Iwasawa2020,Namekata2018}.

\begin{figure}
  \includegraphics[width=0.9\columnwidth]{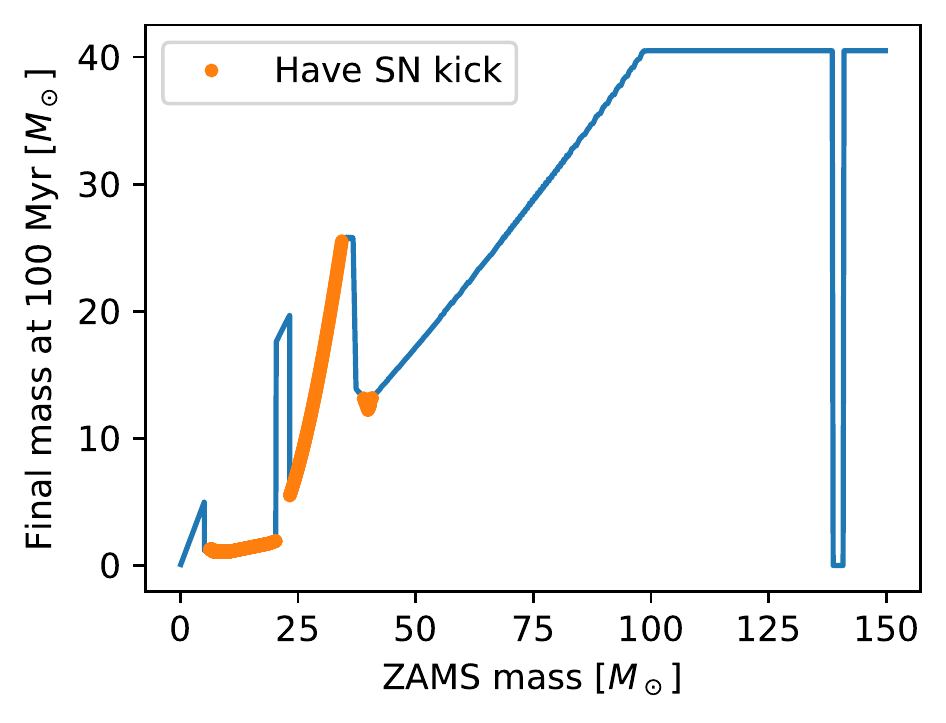}
  \caption{The initial-final mass relation of stars by applying the stellar evolution model from \textsc{sse}. The orange points indicate the remnants natal kick velocity of neutron stars and black holes after the supernova. Due to the fallback treatment, massive BHs have no kick velocity.}
  \label{fig:mm0}
\end{figure}

The single and binary stellar evolution packages, \textsc{sse/bse}, are implemented in \textsc{petar} \citep{Hurley2000,Hurley2002}.
We adopt the updated version of \textsc{sse/bse} from \cite{Banerjee2020b}.
The simulations use the semi-empirical stellar wind prescriptions from \cite{Belczynski2010}, the ``rapid'' supernova model for the remnant formation and material fallback from \cite{Fryer2012}, along with the pulsation pair-instability supernova \citep[PPSN;][]{Belczynski2016}.
In Figure~\ref{fig:mm0}, we show the relation between zero-age main-sequence (ZAMS) masses and final masses of stars from $0.08$ to $150$ $M_\odot$ for $Z=0.001$.
With PPSN, massive stars which have the ZAMS mass above approximately $100 M_\odot$ result in a final mass of BHs with a value of $40.5 M_\odot$.
This has a significant impact on the mass ratio distribution of BBHs, as described in section~\ref{sec:result}.
Meanwhile, the fallback mechanism results in zero natal kick velocity for massive BHs with ZAMS mass $>30 M_\odot$.
They can be retained in the clusters after supernovae, while low mass BHs and most of neutron stars can gain high velocity kicks and immediately escape from the system. 

\subsection{Initial conditions}
\label{sec:init}

To specifically investigate how the shape of IMFs affects the formation and evolution of BBHs in GCs, we develop four models by varying the $\alpha_3$ in the \cite{Kroupa2001} IMF with a multi-component power-law shape:
\begin{equation}
  \begin{aligned}
    &&\xi(m) \propto m^{\alpha_{\mathrm i}} & \\
    \alpha_{\mathrm{1}} & = -1.3,&   0.08 \le m/M_\odot &< 0.50 \\
    \alpha_{\mathrm{2}} & = -2.3,&   0.50 \le m/M_\odot &< 1.00 \\
    \alpha_{\mathrm{3}} &,       &   1.00 \le m/M_\odot &< 150.0.\\ 
  \end{aligned}
  \label{eq:imf}
\end{equation}

The value of $\alpha_3$ and corresponding model names are listed in Table~\ref{tab:init}.
\begin{table}
  \centering
  \caption{The name of $N$-body models with corresponding $\alpha_3$ of IMFs and the number of stars ($N$)}
  \label{tab:init}
  \begin{tabular}{@{}lllll@{}}
  \hline
    Model & A1.5 & A1.7 & A2.0 & A2.3\\
    \hline
    $\alpha_3$ & -1.5 & -1.7 & -2.0 & -2.3\\
    N & 182306 & 312605 & 581582 & 854625 \\
    \hline
  \end{tabular}
\end{table}
A2.3 has the canonical value of $\alpha_3$, while all the others correspond to different degrees of top-heavy IMFs.
The maximum value of $\alpha_3$ is chosen to be 1.5.
The minimum and the maximum masses of stars in our models are $0.08$ and $150$ $M_\odot$, respectively.
For the metallicity, which determines the stellar evolution (mass loss) of stars, we adopt a typical value for GCs, that is, $Z=0.001$.

For all models, we adopt the Plummer profile to set up the positions and velocities of individual stars.
We fix the initial total mass to $5\times10^5 M_\odot$ and the initial half-mass radius, $r_{h,0}$, to 2 pc.
Thus, our models have a mass and density as high as those typically observed in GCs.
In fact, the density of our model is higher than that of the previous largest DRAGON GC model \citep{Wang2016}.
Thus, our simulations are still time-consuming even after using the \textsc{petar}.

To begin with, we assume no primordial binaries being present in these models.
This simplifies the discussion since we can purely focus on the dynamical formation of BBHs.
Meanwhile, it is easier to perform the $N$-body simulations. We do not apply the galactic potential either, but we remove unbounded stars when they reach more than 200~pc far from the cluster centre.

\section{Results}
\label{sec:result}

We do not intend to create realistic GC models with primordial binaries and galactic tidal field, but focus on the theoretical studies of the BBHs in GCs.
Thus, we only evolve our models up to $3$~Gyr instead of Hubble time, using the available computing resource.
However, this already covers approximately $10$ initial half-mass relaxation times of the systems, and thus, it is sufficient for analysis.

\subsection{BBH mergers}

In our simulations, we do not model BBH mergers via the effect of general relativity.
Instead, we detect all BBHs that are potential GW mergers in the post-process using the snapshots of the simulations.
We apply the following two steps to obtain the BBH mergers.

Firstly, we estimate the merger timescale for each binary by using the formula provided by \cite{Peters1964}:
\begin{equation}
  \label{eq:tgw}
  \begin{aligned}
    \tgw & = \frac{12}{19}\frac{c_0^4}{\beta}  \int_0^{e_0} {\frac{e^{29/19}[1+(121/304)e^2]^{1181/2299}}{(1-e^2)^{3/2}} \diff e} \\
    c_0 & = \frac{a_0 (1-e_0^2) [1 + (121/304) e_0^2]^{870/2299}}{e_0^{12/19}}\\
    \beta & = \frac{64}{5}\frac{G^3 m_1 m_2 (m_1 + m_2) }{c^5}.\\
  \end{aligned}
\end{equation}
By calculating $\tgw$ of BBHs in the snapshots with the time interval of $0.25-1$ Myr, we select the merger candidates which have the actual (delayed) merging time $\tmerge \equiv \tgw + t < 12 $Gyr, where $t$ is the physical time of the cluster when $\tgw$ is calculated.

However, these candidates may not actually merge because BBHs with high eccentricities can be perturbed by their surrounding stars, if they are still in the cluster. 
The orbits of BBHs can also dramatically change after a strong few-body interaction. 
If the eccentricity decreases, $\tgw$ can significantly increases.
Thus, in the second step, we look at the evolution of each candidate and check whether they escape from the system or merge before their $\tgw$ increases.
Some BBHs can suffer exchange of members after interaction with other BHs or BBHs (hereafter referred to as exchanged BBH).
For example, after an interaction between a BBH (BH1, BH2) and a single BH3, one member (BH2) in the BBH may be exchanged to BH3.   
In such a case, we first check whether BH1 and BH2 can merge before exchange.
If not, we check whether the new BBH (BH1, BH3) can merge.
If the merger occurs in any case, we consider it as one merger event. 

The numbers of candidates, the candidates without and with exchanged BBHs, and confirmed mergers for each model are shown in Table~\ref{tab:bbhevent}.
We can notice that the number of candidates does not monotonically depends on the $\alpha_3$ of IMFs. 
The more top-heavy IMF is, less the $\ncand$ tend to be, except for that of the A1.5 model.
The A1.5 model has more $\ncand$ than that of the A1.7.
Besides, the number of exchanged BBHs is large in A2.0 and A2.3 models. 
We will explain the reason in Section~\ref{sec:se}.

\begin{table}
  \centering
  \caption{The number of BBH merger candidates ($\ncand$), the candidates without ($\ncandsin$) and with ($\ncandex$) exchanged BBHs mergers inside GCs ($\nin$) and escaped mergers ($\nout$) up to the 3~Gyr evolution of GCs. The exchanged BBHs are counted multiple times in $\ncand$.}
  \label{tab:bbhevent}
  \begin{tabular}{@{}lllll@{}}
  \hline
    Model & A1.5 & A1.7 & A2.0 & A2.3\\
    \hline 
    $\ncand$ & 20 & 10 & 30 & 43 \\
    $\ncandsin$ & 15 & 8 & 18 & 19 \\
    $\ncandex$ & 2 & 1 & 5 & 11 \\
    $\nin$ & 0  & 1  & 3  & 3 \\
    $\nout$ & 3 & 3  & 3  & 4 \\
    \hline
  \end{tabular}
\end{table}

\begin{figure}
  \includegraphics[width=0.95\columnwidth]{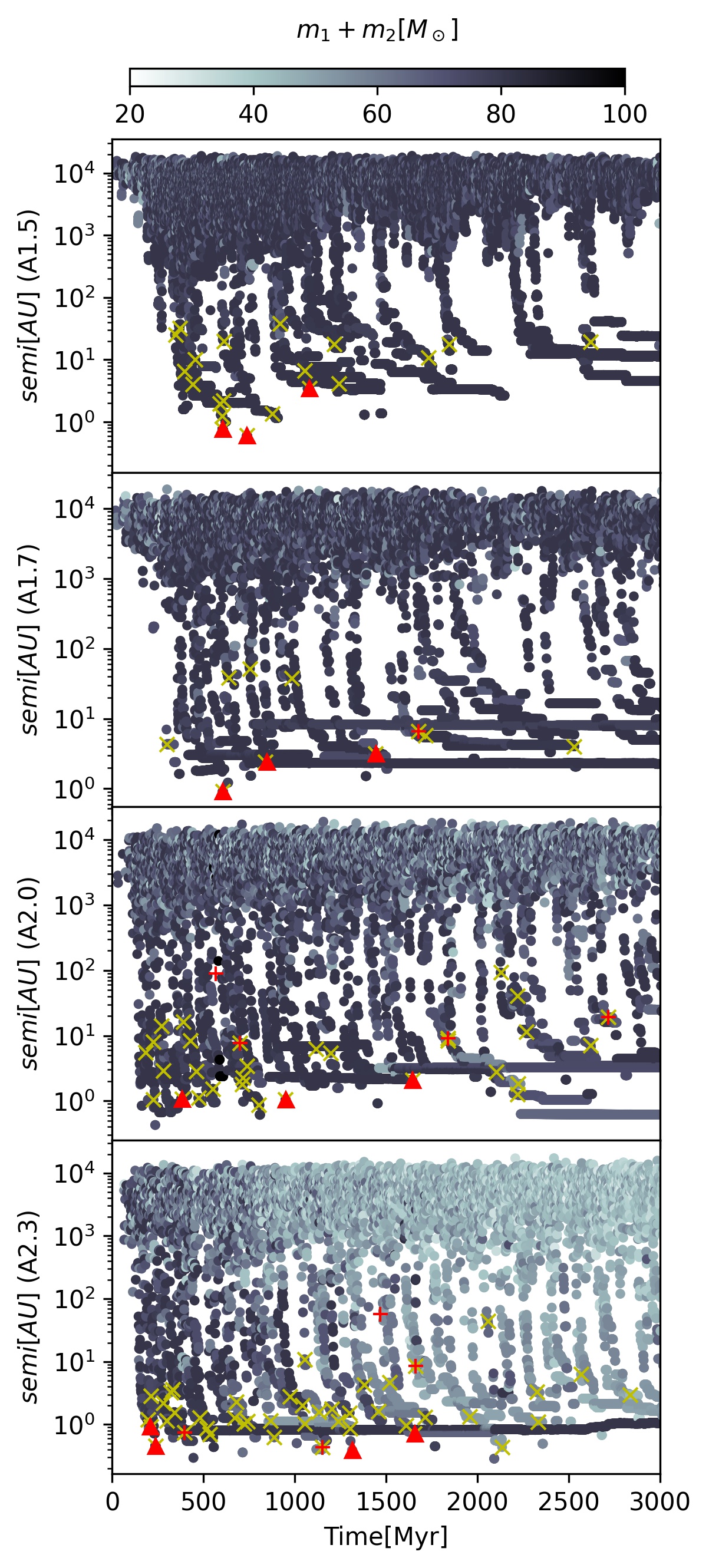}
  \caption{The evolution of semi-major axes of all BBHs in the four models.
    The gray scale indicates the total masses of BBHs.
    The yellow ``x'' represents the merger candidates.
    The red triangles and red crosses represents escaped mergers and mergers inside GCs.
  }
  \label{fig:bbhat}
\end{figure}

Figure~\ref{fig:bbhat} shows the evolution of the semi-major axis ($a$) of all BBHs detected in the snapshots of our simulations.
The candidates, inner and escaped BBH mergers are shown together.
In this figure, we can identify two types of BBHs, the soft and hard ones separated by $a \approx 10^3$ AU.
As explained by the Heggie-Hills law \citep{Heggie1975,Hills1975}, soft binaries are easily disrupted by close encounters, but hard binaries become tighter.
This can be seen in Figure~\ref{fig:bbhat}.
Soft BBHs randomly appear and vanish, while hard binaries continuously evolve harder, following a clear trace of decreasing $a$.
Typically, only one or two hard BBHs appear at one time because binary-binary encounters tend to break softer binaries. 
Once their $a$ becomes small enough, a strong encounter can eject them out of the GCs, and they disappear.
Then, the new hard BBHs are formed, and the same process is repeated.

\begin{figure}
  \includegraphics[width=0.9\columnwidth]{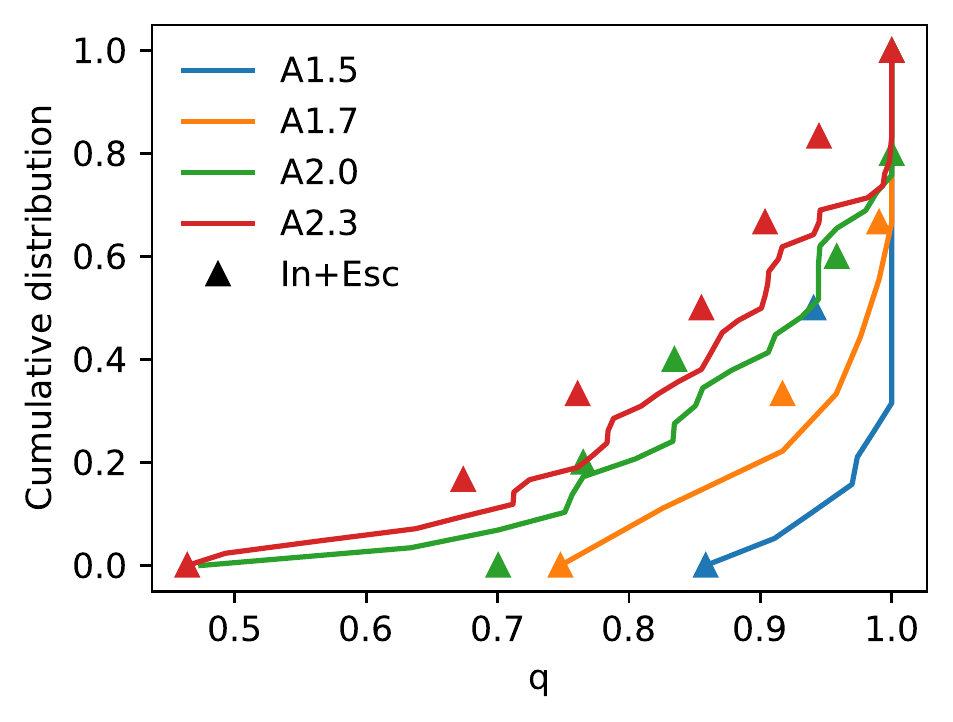}
  \caption{The cumulative distribution of mass ratio $q$ of two components in BBH candidates (lines) and confirmed (triangles) mergers.}
  \label{fig:q}
\end{figure}

From the A2.3 model, we can identify a clear trend that BBHs with large masses prefer to form first.
Once they escape, BBHs with lower masses form and escape one by one.
This feature results in a clear difference in the mass ratio distribution of BBH components depending on IMFs, as shown in Figure~\ref{fig:q}.
As more top-heavy IMFs have larger fractions of stars with the ZAMS masses above $100 M_\odot$, more equal-mass BHs of $40.5 M_\odot$ form as a result of the PPSN.
These most massive BHs form binaries first.
Thus, the BBH mergers in top-heavy IMFs tend to have $q$ closer to unity.

Another major feature shown in Figure~\ref{fig:bbhat} is the increase of minimum semi-major axis (maximum binding energy), especially in the A1.5 model.
As a result, it becomes more difficult to form tight BBHs in the later evolution of GCs.
This is reflected on the $\ncand$ as a function of time.
Most of the candidates form in the first 1000 Myr in the A1.5 model, and the formation rate significantly decreases after 2000 Myr.
In Section~\ref{sec:bhheating}, we explain why the minimum semi-major axis increases by analyzing the escape velocities of GCs.

Meanwhile, the formation rate of tight BBHs is higher when IMF is more top-light.
This also explains the larger $\ncand$. 

\begin{figure}
  \includegraphics[width=0.9\columnwidth]{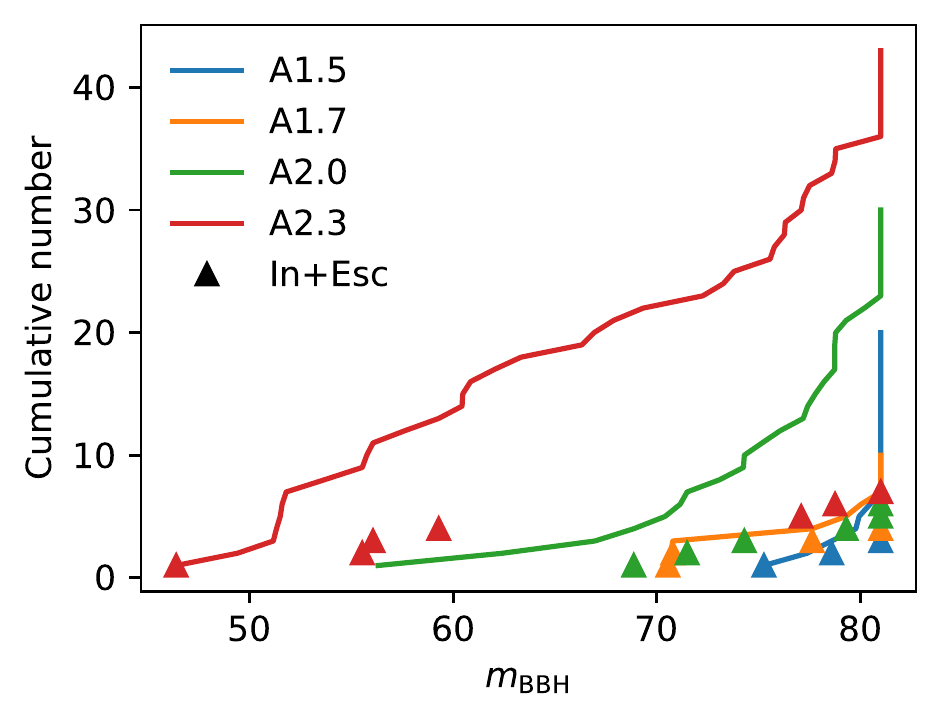}
  \caption{The cumulative number of BBHs candidates (solid curves) and confirmed ones (triangles) vs. the masses of BBHs.}
  \label{fig:mbbh}
\end{figure}

In Figure~\ref{fig:mbbh}, we show the cumulative distribution of the masses of BBH merger candidates and confirmed ones. 
There are only a few confirmed mergers with $\mbbh<60 M_\odot$ in the A2.3 model.
In A1.5 models, however, most mergers have $\mbbh=81 M_\odot$ with the mass of each component being $\mbbh=40.5 M_\odot$ due to the PPSN.
Thus, clusters with a more top-heavy IMF form more massive mergers.

\begin{figure}
  \includegraphics[width=0.95\columnwidth]{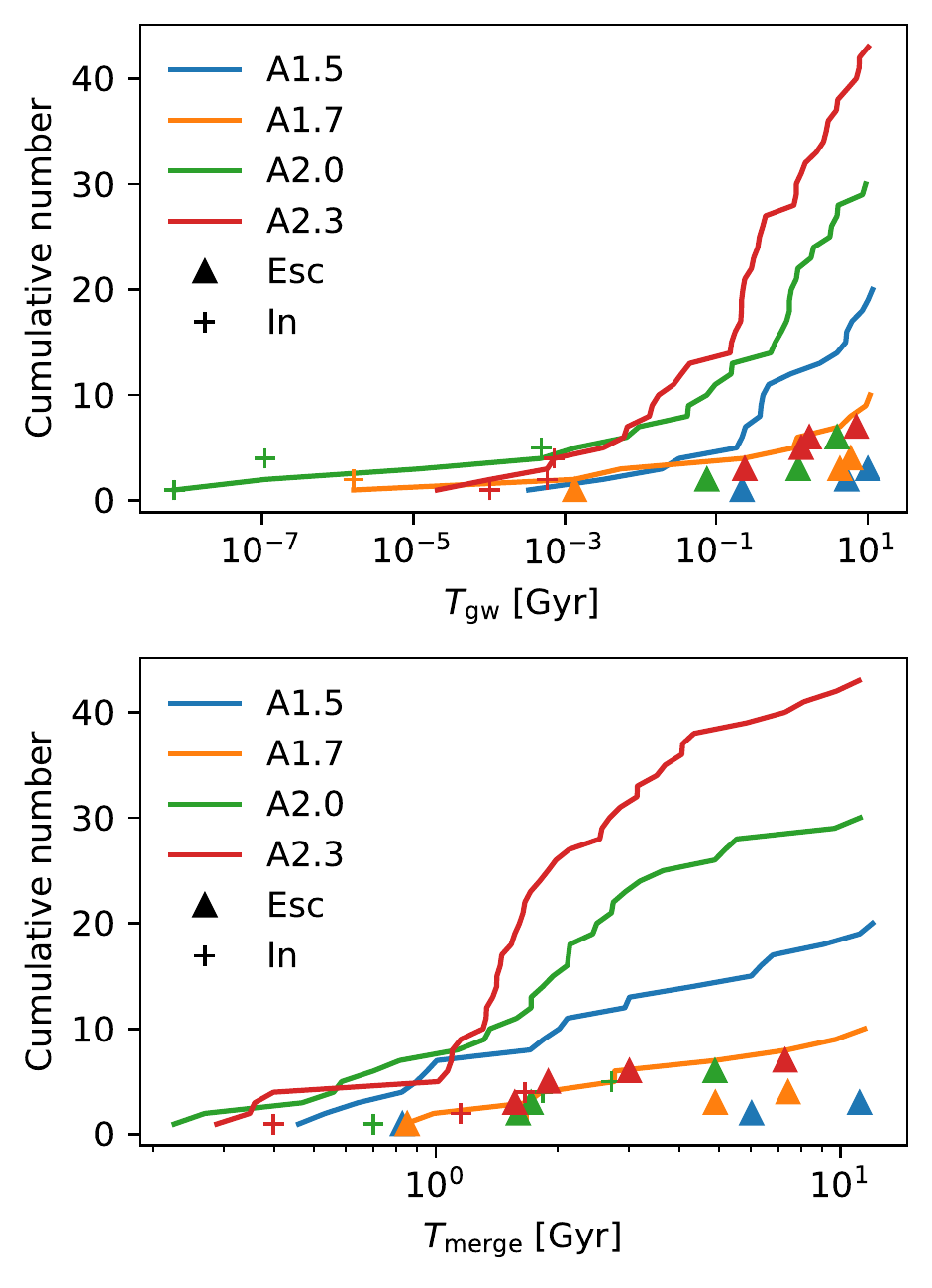}
  \caption{The cumulative number of BBH merger candidates (solid curves), mergers inside GCs (crosses) and escaped mergers (triangles) vs. $\tgw$ (upper panel) and $\tmerge$ (lower panel).}
  \label{fig:cdtgw}
\end{figure}

Here, the analysis includes all BBH mergers. 
However, the mergers that occurred at an early time are not observable.
In Figure~\ref{fig:cdtgw}, we show the cumulative distribution of $\tgw$ and $\tmerge$ for BBH merger candidates and confirmed ones for all models.
While mergers inside GCs have a short merger time ($\tgw<1$~Myr and $\tmerge<3$~Gyr), most escaped mergers have a relatively long merger time ($\tgw>1$~Gyr and $\tmerge=1-10$~Gyr).
Therefore, the escaped mergers from all models with different IMFs can be detected today.

In our models with the age of 3 Gyr, there are only a few BBH merger candidates with redshift $< 1$ (within approximately 4 Gyr of lookback time).
Thus, we cannot provide the statistical analysis of the model directly.
However, if all mergers after 3 Gyr are included, we expect that the main trend, i.e., top-heavy IMFs result in more massive and more $q \sim 1$ mergers, becomes stronger.
In the later evolution (after 3 Gyr), all models with different IMFs will have more low-mass BBH mergers.
However, the top-heavy model has a lower escape velocity, and the BBH merger rate is lower than that of the top-light model.
Thus, more low-mass BBH mergers will appear in the top-light model.

\subsection{Effect of stellar evolution on dynamics}
\label{sec:se}

\begin{figure}
  \includegraphics[width=0.95\columnwidth]{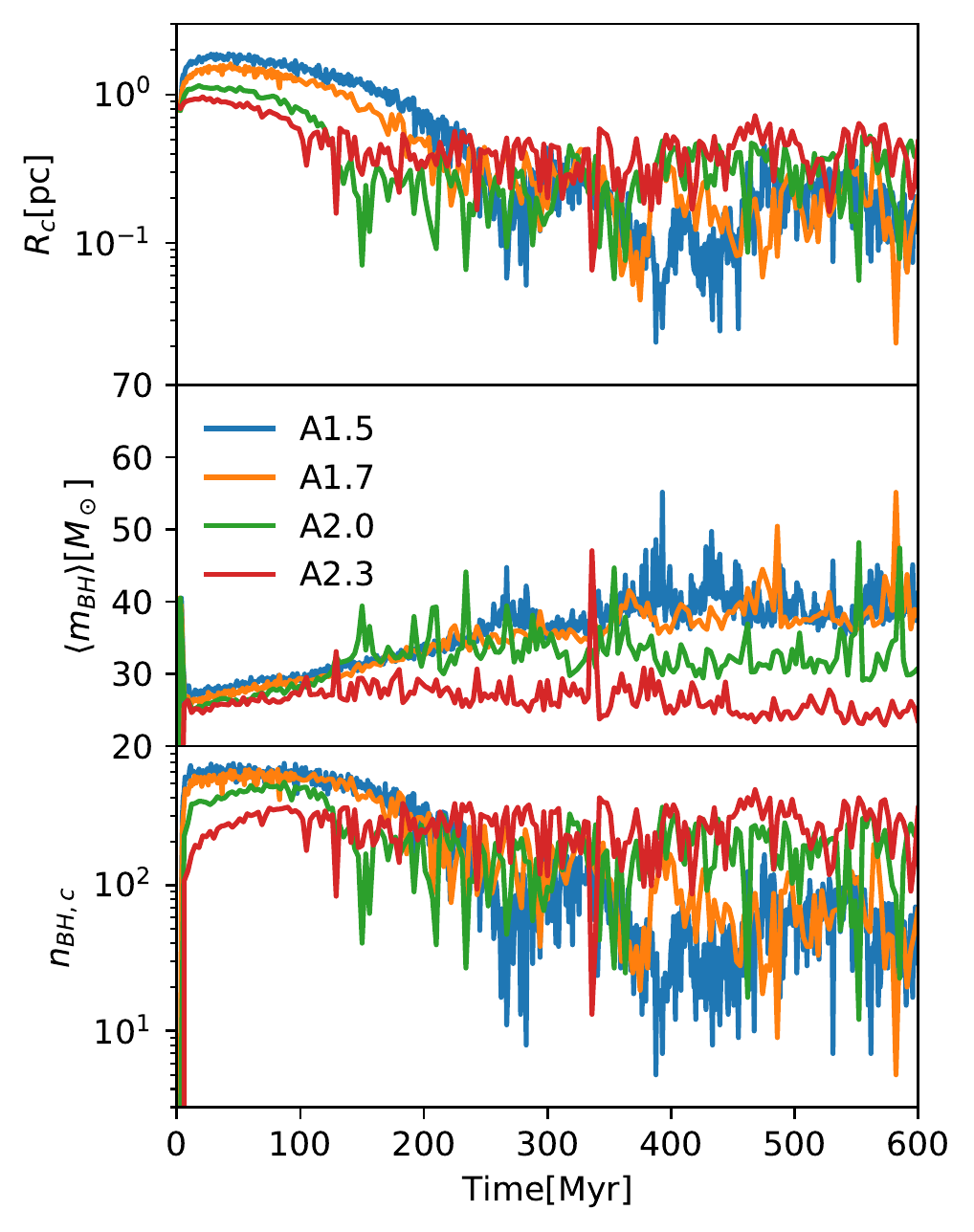}
  \caption{The evolution of core radii ($\rc$) defined by Equation~\ref{eq:rc}; averaged masses ($\mbhc$) and numbers of BHs ($\nbhc$) inside $\rc$. BBHs are counted as single objects.}
  \label{fig:rct}
\end{figure}

The stellar evolution, especially the wind mass loss of massive stars in the first $100$~Myr, has a significant impact on the later evolution of GCs \citep[e.g.][]{Trani2014}.
In particular, the mass loss with a top-heavy IMF is more intense due to the larger fraction of massive stars.
This significantly affects the central density of GCs.
To investigate this, we calculate the evolution of the core radii ($\rc$) defined by \cite{Casertano1985} as
\begin{equation}
    \rc = \sqrt{\frac{\sum_i \rho_i^2 r_i^2}{\sum_i \rho_i^2}},  
    \label{eq:rc}
\end{equation}
where $\rho_i$ is the local density of object $i$ estimated by counting 6 nearest neighbors, and $r_i$ is the distance to the center of the system.
As shown in Figure~\ref{fig:rct}, $\rc$ in the first 100~Myr becomes larger with more top-heavy IMFs.
When most massive stars evolve to compact remnants, the stellar-wind mass loss becomes weak.
Then, $\rc$ starts to shrink, and finally, the core collapse occurs.
It must be noted here that the core is dominated by BHs after mass segregation\footnote{The theoretical $\rc$ discussed in this work includes all objects in the system. This is not the same as the core radius defined in observation, which is measured by fitting the surface brightness profile. The core collapse actually occurs in the BH subsystem while the observed core radius is much more extended.}.
Figure~\ref{fig:rct} shows that the A1.5 model contracts the most during the core collapse, probably due to the larger masses of the BHs, and the longer distance of sinking.
This causes the formation of the densest core between $350-450$ Myr compared to those of the other models.
Such a high density might be the reason why the A1.5 model has more BBH merger candidates compared to that of the A1.7 model.
Figure~\ref{fig:bbhat} shows that many of the BBH merger candidates appear around 400~Myr in the A1.5 model.
This is consistent with the feature of core collapse.

\subsection{Expansion due to BH heating}
\label{sec:bhheating}

\begin{figure}
  \includegraphics[width=0.95\columnwidth]{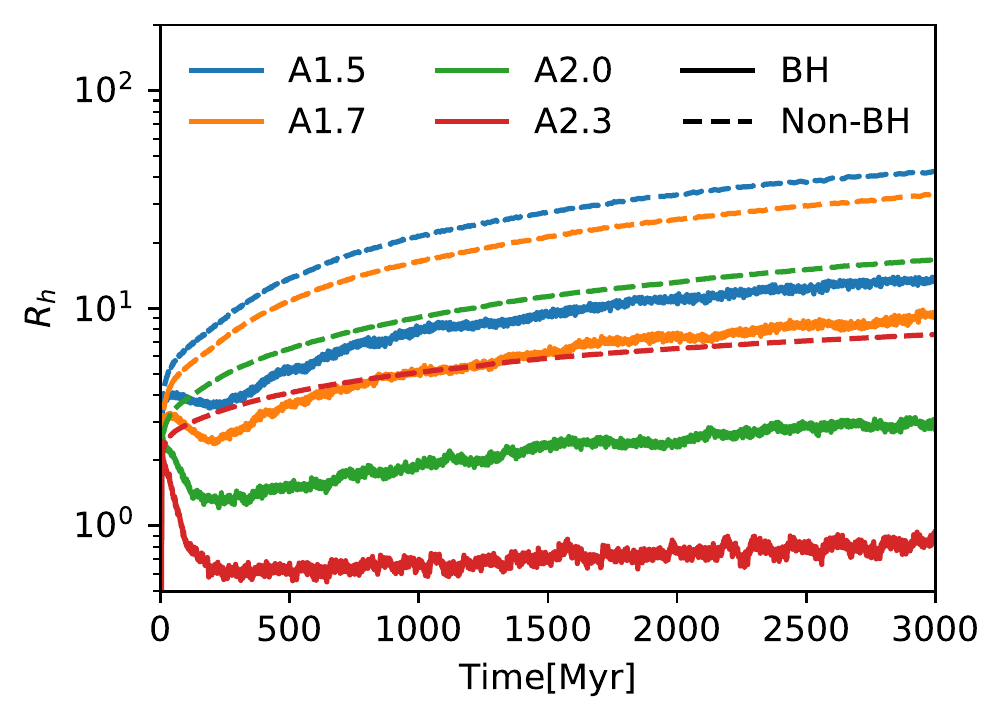}
  \caption{The evolution of half-mass radii of non-BH objects ($\rhnobh$; dashed curves) and BHs ($\rhbh$; solid curves).}
  \label{fig:rht}
\end{figure}

Due to the larger masses of BHs compared to those of stars, BHs have a strong impact on the long-term evolution of star cluster \citep{Breen2013}.
In star clusters including BHs, massive BHs sink into the cluster center due to the dynamical friction (mass segregation) and form a dense subsystem.
Thereafter, the core collapse of the BH subsystem occurs and drives the formation of BBHs, which heats the systems via few-body interactions.
As a result of the energy-balanced evolution, the halo composed of light stars expands.
When more BHs exist, the expansion is faster \citep[see also in][]{Giersz2019,Wang2020b}.
This can be identified in Figure~\ref{fig:rht}, in which the evolution of the half-mass radii of non-BH objects ($\rhnobh$) and BHs ($\rhbh$) are compared for all models.

In the first 300 Myr, $\rhbh$ decreases due to the mass segregation, and later, it increases due to BH heating. Meanwhile, $\rhnobh$ always increases.
The stronger stellar-wind mass loss in more top-heavy models results in larger $\rhbh$ and $\rhnobh$ in a short time ($<100$ Myr).
This is similar to the evolution of the core radii shown in Figure~\ref{fig:rct}.
As a result of stronger BH heating during the long-term evolution after 300 Myr, $\rhbh$ and $\rhnobh$ increase faster in the more top-heavy models.

According to Heggie-Hill law, the boundary of the semi-major axis between soft and hard BBHs depend on the masses of BHs and local velocity dispersion:
\begin{equation}
  a_{\mathrm{s/h}} = \frac{G m_1 m_2}{\langle m \rangle \sigma^2} 
  \label{eq:hhlaw}
\end{equation}

\begin{figure*}
  \includegraphics[width=0.8\textwidth]{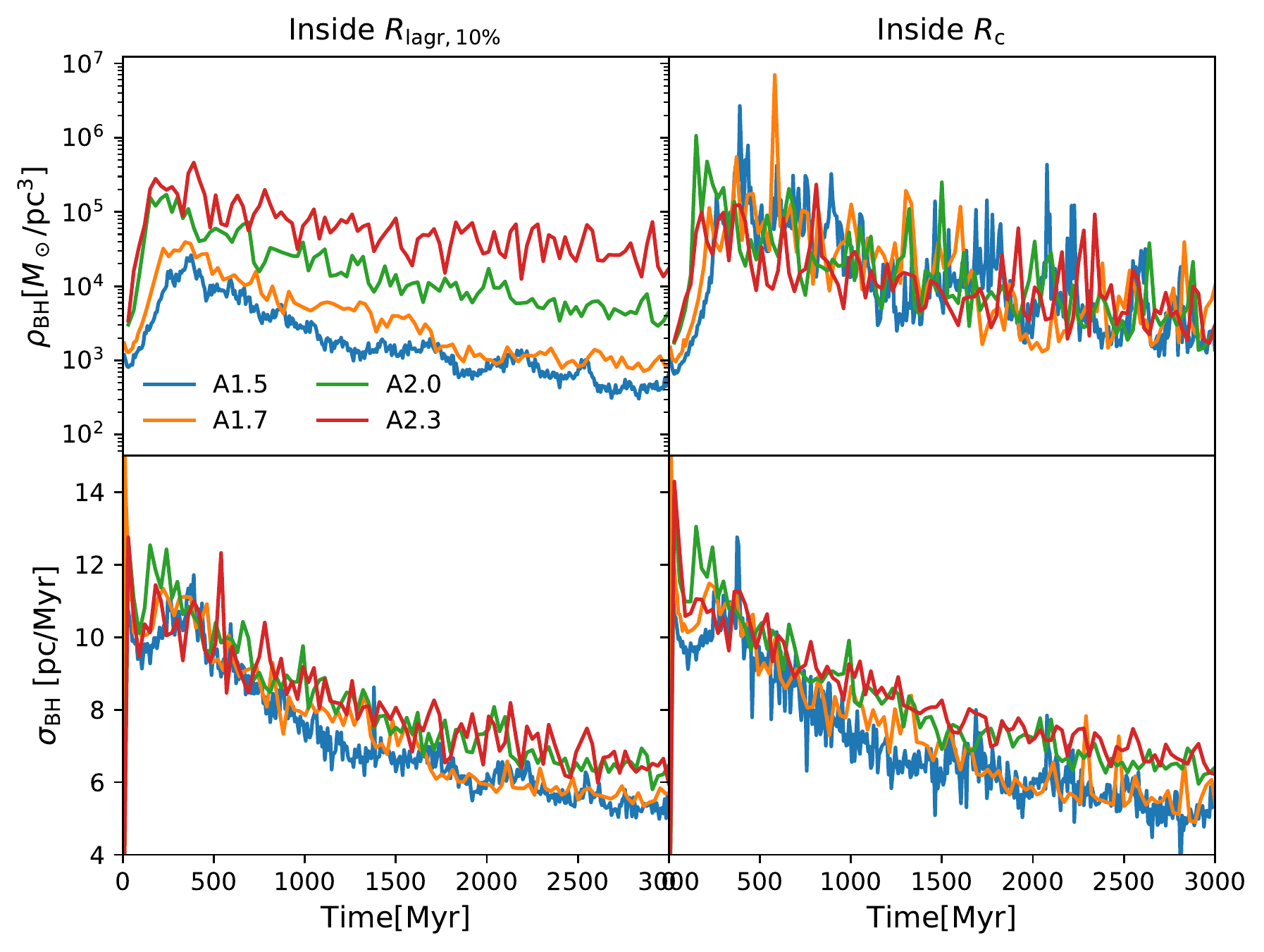}
  \caption{The evolution of density of BHs ($\rhobh$) and 3D velocity dispersion of BHs ($\sigmabh$) within $10\%$ Lagrangian radii ($\rlagrcbh$) and $\rc$}
  \label{fig:rhot}
\end{figure*}

We check the density ($\rhobh$) and 3D velocity dispersion ($\sigmabh$) of the BHs within $10\%$ Lagrangian radii of all BHs and $\rc$ of the BH subsystem.
The results are shown in Figure~\ref{fig:rhot}.
We find that $\rhobh$ inside $\rlagrcbh$ varies as the slope of the IMF changes, i.e.,  
models with more top-heavy IMFs have lower central densities.
This is consistent with the behaviour of $\rhbh$ (see Figure~\ref{fig:rct}).
In contrast, $\rhobh$ inside $\rc$ is very similar for all models, different from the behaviour on a larger distance scale.

We can divide the models into two groups, A1.5/A1.7 and A2.0/A2.3, based on the values of $\sigmabh$ within $\rlagrcbh$ or $\rc$. There is a gap of $\sim 2$ pc/Myr in $\sigmabh$ between the two groups.
Models with more top-heavy IMFs have smaller $\sigmabh$ values, and thus, their $a_{\mathrm{s/h}}$ are larger as explained by Equation~\ref{eq:hhlaw}.
This is consistent with the hard-soft boundaries shown in Figure~\ref{fig:bbhat}.

As clusters expand, $\sigmabh$ continues to decrease for all models, and $a_{\mathrm{s/h}}$ increases.
This is identified in the A1.5 and A1.7 models in Figure~\ref{fig:bbhat}.
In the A2.0 and A2.3 models, however, we can clearly see that the masses of BBHs decrease with time.
This balances the effect of decreasing $\sigmabh$, and thus, the change in $a_{\mathrm{s/h}}$ is not obvious for these two models.

The key quantity that controls the minimum semi-major axis of binaries is the central escape velocity of star clusters.
As the semi-major axis of a hard binary shrinks, the binary can gain larger kinetic energy (larger kick to the center-of-mass velocity) after a strong encounter.
This process finally causes the ejection of the binary from the center of the cluster after a strong encounter.
If the ejection velocity is below the escape velocity, the dynamical friction brings it back to the center.
The binary can continue to encounter with others and become harder, until the next strong encounter ejects it again.
Therefore, the escape velocity of star clusters limits the minimum semi-major axis of hard binaries that can be reached by few-body encounters.

Based on the mechanism described above, we can estimate GW merger timescale ($\tgw$) of an ejected BBH, which should be an  indicator for $\tgw$ of BBHs in a star cluster. 
According to Equation~\ref{eq:tgw} \citep{Peters1964}, $\tgw \propto a_{\rm esc}^4/m_1^3$, where $a_{\rm
    esc}$ is the semi-major axis of an ejected BBH, and we assume $m_2
  = m_1$ for the BBH. Since an ejected BBH has an internal velocity of
  $\sim \vesc$, $a_{\rm esc} = Gm_1/(2\vesc^2)$, and $\tgw \propto
  m_1/\vesc^8$. Eventually, an ejected BBH has larger $\tgw$ when it contains heavier BHs, and when it is ejected from a star cluster with a
  smaller escape velocity.

\begin{figure}
  \includegraphics[width=0.9\columnwidth]{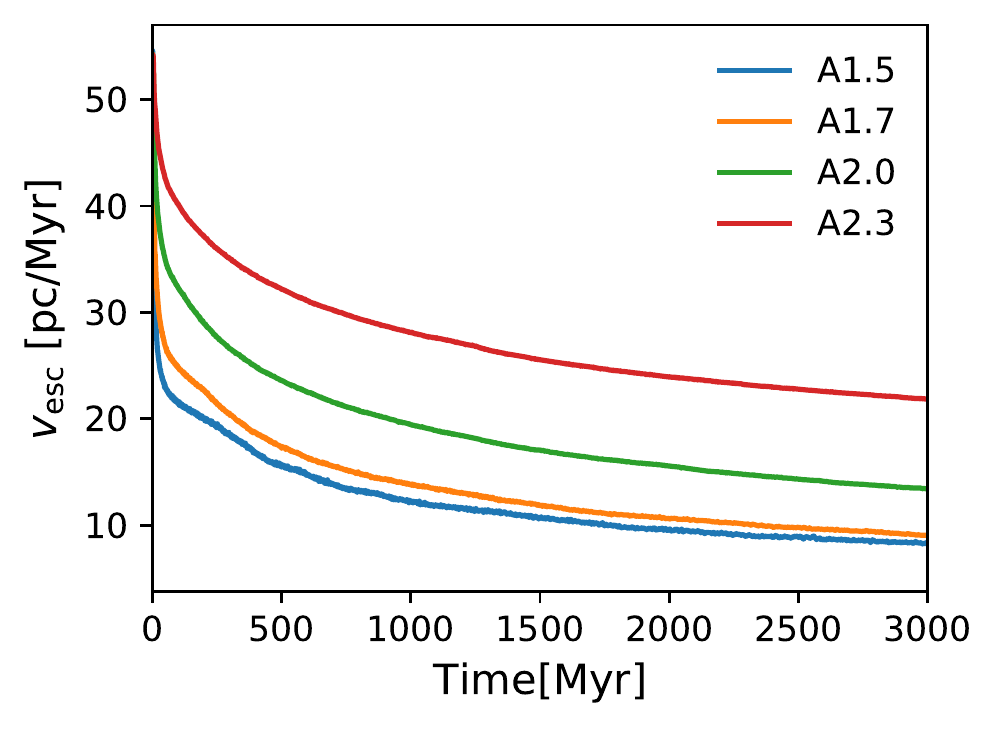}
  \caption{The evolution of the central escape velocity of the models}
  \label{fig:vesct}
\end{figure}

In Figure~\ref{fig:vesct}, we estimate the central escape velocity of
the Plummer model,
\begin{equation}
  \vesc = \sqrt{2\times 1.305 G \frac{M}{\rh}},
  \label{eq:esc}
\end{equation}
where $M$ is the total mass of the system. Due to the larger $\rh$ in models with top-heavy IMFs, their $\vesc$ is smaller. Moreover, their BHs are heavier in models with more top-heavy IMFs. Therefore, it is difficult to form BBH mergers there, although the number of BHs is much larger. 
Meanwhile, $\vesc$ decreases, as the cluster expands. 
This explains why the minimum semi-major axis increases as shown in Figure.~\ref{fig:bbhat}, especially for the A1.5 model, because the masses of BBHs maintain a similar value up to 3~Gyr. 
In the case of the A2.3 model, the masses of BBHs decrease, while the minimum semi-major axis does not show a strong increasing trend.

Since the models with top-heavy IMFs have a stronger expansion and maintain heavier BHs inside the clusters, the efficiency of forming BBH mergers decreases faster.  
Therefore, with the same initial total mass and size, the BBH merger rate is actually lower in the GCs with top-heavy IMFs. 
If the cluster with a top-heavy IMF was under a strong tidal field, the decrease of $\vesc$ would be faster, and therefore, it would be more difficult for the cluster to generate merging BBHs.  
However, if the cluster was initially denser and more massive, it could survive up to Hubble time and continue to generate merging BBHs all the time. In such a case, the cluster with a top-heavy IMF contains a large number of BHs, and therefore, the total contribution of BBH mergers can be significantly larger than that in a GC with the same mass and size but a top-light IMF. 
Such `dark clusters' have been studied using Monte-Carlo simulations \citep{Weatherford2021}.

\subsection{Mass loss}
\label{sec:massloss}

\begin{figure}
  \includegraphics[width=0.9\columnwidth]{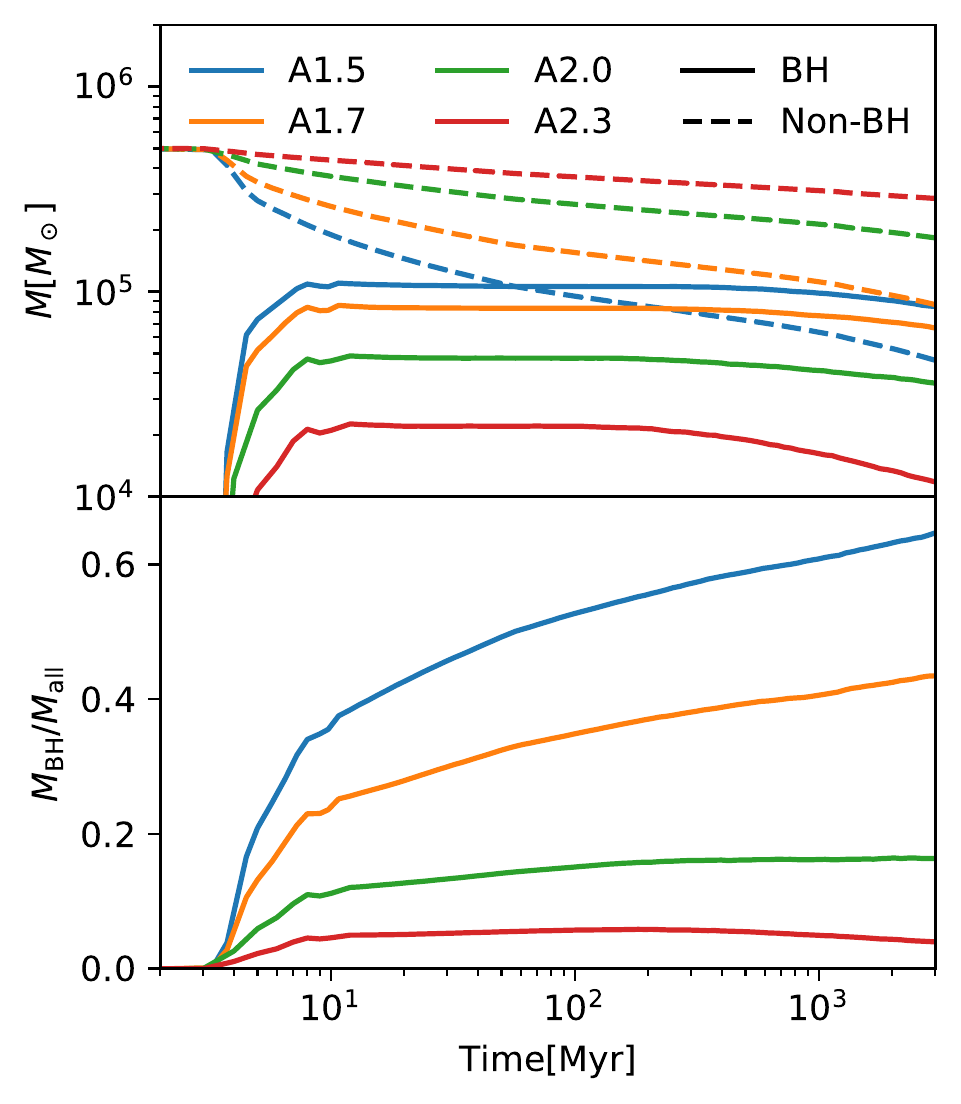}
  \caption{The evolution of total masses of BH and non-BH components (upper panel) and mass ratio between these two (lower panel).}
  \label{fig:mt}
\end{figure}

The stellar wind and BH heating drive the mass loss of star clusters in the early and later phases, respectively.
Figure~\ref{fig:mt} (upper panel) shows the time evolution of the total masses of BH and non-BH components in the four models.
After 100~Myr, the A1.5 model loses about $60\%$ of the initial mass due to stellar-wind mass loss, while the A2.3 model loses only $30\%$.
Therefore, the impact of the stellar evolution is more significant in star clusters with more top-heavy IMFs.
Such strong mass loss and the subsequent BH heating drive the fast expansion of the system as shown in Figure~\ref{fig:rct} and \ref{fig:rht}.
This is also reported in previous studies \citep[e.g.][]{Chernoff1990,Banerjee2011,Chatterjee2017,Giersz2019}.
Besides, after the core collapse of BH subsystems, the few-body interactions between BBHs and others start to eject BHs from the GCs, and thus, $\Mbh$ starts to decrease after approximately 300 Myr.
Since our models do not have a tidal field but simply remove unbound stars above 200 pc, the A1.5 model still survives and maintains approximately $34\%$ initial mass at 3~Gyr.
In the realistic condition, where the galactic tidal field plays a role, the clusters with top-heavy IMFs tend to dissolute much faster than those in our current models \citep{Wang2020b}.

\cite{Breen2013} and \cite{Wang2020b} have found that the mass fraction of BHs in the system ($\Mbh/M$) evolves depending on the initial fraction and the tidal field.
Since BHs are centrally concentrated, they are not directly affected by the tidal field.
Thus, the tidal evaporation of BHs can be neglected.
As mentioned in Section~\ref{sec:bhheating}, strong close encounters between hard BBHs and intruders can eject BHs from the cluster.
This is the major scenario that causes the mass loss of BHs.

Meanwhile, light stars in the halo are truncated by the tidal field. When BH heating occurs, this process is accelerated \citep{Breen2013,Giersz2019,Wang2020b}.
In the lower panel of Figure~\ref{fig:mt}, we can identify the two different evolution trends of $\Mbh/M$.
In the A2.3 model, $\Mbh/M$ decreases with time, and finally, most of the BHs will be ejected from the clusters. The core collapse of light stars will occur, and a GC with a dense core will appear in the observation.

In contrast, $\Mbh/M$ in the A1.5 and A1.7 models increases with time. This means that light stars will finally evaporate from such clusters and that dark clusters will form.
In a strong tidal field, the mass loss of light stars is faster, and such this process can be accelerated. The A2.0 model is in the transition region.

\begin{figure}
  \includegraphics[width=0.9\columnwidth]{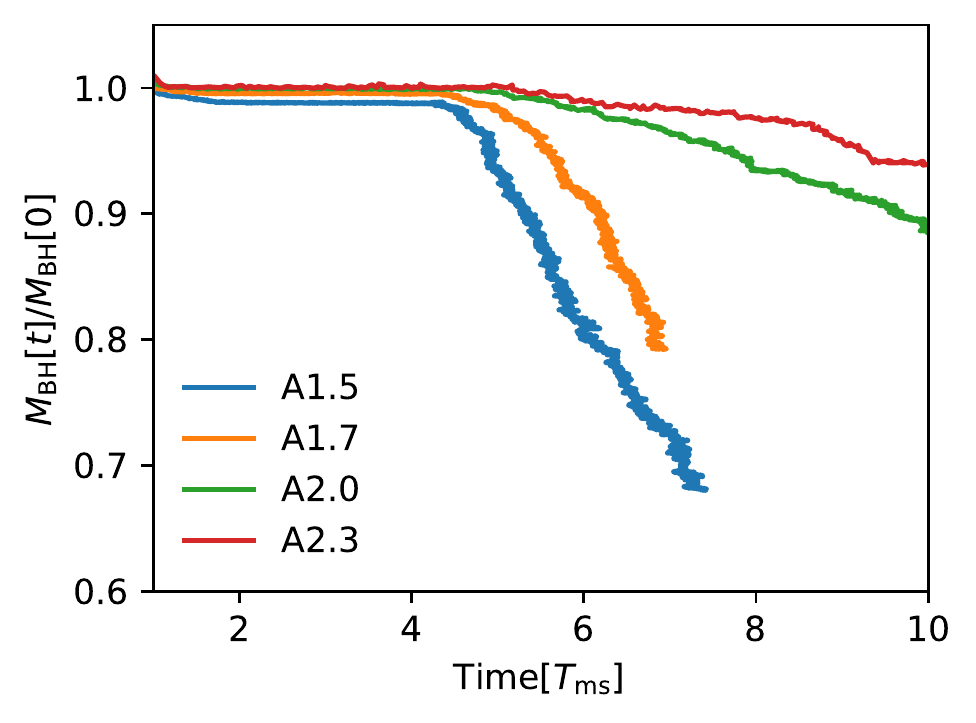}
  \caption{The evolution of total mass of BHs, $\Mbh$($t$). The mass is normalized by using the value at $100$~Myr. Time is in the unit of $\tms$ (eq. \ref{eq:tms}).}
  \label{fig:mtms}
\end{figure}

With a two-component simplified model, \cite{Wang2020b} found that the mass loss rate of BHs, $\Mbh(t)/\Mbh(0)$, simply depends on the mass segregation time of heavy components ($\tms$) in isolated clusters, but the dependence becomes complex if a tidal field exists.
In our models with stellar evolution and IMFs, even without a tidal field, $\Mbh(t)/\Mbh(0)$ does not simply depend on $\tms$ as shown in Figure~\ref{fig:mtms}.
In a two-component model, the definition of $\tms$ is as follows:
\begin{equation}
  \tms = \frac{m_1}{m_2}\trha,
  \label{eq:tms}
\end{equation}
where the relaxation time of the ight component has the form \citep{Spitzer1971}
\begin{equation}
  \trha = 0.138 \frac{N^{1/2} \rh^{3/2}}{m_1^{1/2} G^{1/2} \ln \Lambda}.
  \label{eq:trh}
\end{equation}
However, with a mass function, the precise definition of $\tms$ is difficult to determine.
Here, we use the averaged mass of non-BH and BH components as a replacement of $m_1$ and $m_2$ in Equation~\ref{eq:tms} and \ref{eq:trh}.
Theoretical studies have shown that when multiple components exist, the diffusion coefficients need to be properly averaged to obtain the correct relaxation time \citep{Spitzer1971,Antonini2019b,Wang2020b}.
Thus, using averaged mass is not accurate to calculate $\trha$.
In particular, since the mass functions are different among the four models, we probably need to introduce a correction factor $\psi$ to $\trha$, similar to the case in the two-component models done in \cite{Wang2020b}.
Moreover, the stellar evolution introduces another complexity.
Thus, it is reasonable to see that $\Mbh(t)/\Mbh(0)$ does not simply depend on $\tms$.
Instead, Figure~\ref{fig:mtms} shows that the mass loss rate of BHs is faster when IMF is more top-heavy.

\subsection{Energy-balanced evolution}

\cite{Breen2013} established a theory to describe the long-term evolution of star clusters with BH subsystems.
The key idea is based on the energy-balanced evolution of star clusters \citep{Henon1961,Henon1975}.
After the BH subsystem forms in the center of star clusters due to the mass segregation, BHs drive the binary heating process and provide the energy to support the whole system. 
The H{\'e}non's principle suggests that the energy flux from the center (BH subsystem) should balance the one required by the global system.
This can be described by \citep[Eq.~1 in ][]{Breen2013}
\begin{equation}
  \frac{E}{\trh} \approx k \frac{\Eb}{\trhb} ,
  \label{eq:eb}
\end{equation}
where $\Eb$ and $E$ are the total energy of the BH subsystem and the global system respectively; $\trh$ and $\trhb$ are the two-body relaxation times measured at $\rh$ and $\rhbh$, respectively.

This relation constrains the behaviour of BHs, i.e., the density of the BH subsystem.
The formation rate of BBHs and the escape rate of BHs are controlled by the global system (light stars), and not the BH subsystem itself. 
By extending the \cite{Breen2013} theory to top-heavy IMFs, \cite{Wang2020b} found that when a large fraction of BHs exists, the energy balance is different from the description of Equation~\ref{eq:eb}.
To properly measure the energy balance between the central BH subsystem and the global system, the correction factor $\psi$ is required to define the relaxation time more appropriately (as discussed in Section~\ref{sec:massloss}).
In \citet{Wang2020b}, $\psi$ is defined as
\begin{equation}
      \psi = \frac{\sum_k \nk \mk^2/\vk}{\n \m^2 /\vave},
    \label{eq:psi}
\end{equation}
where $\nk$, $\mk$ and $\vk$ are the number density, the mass of one object and the mean square velocity of the component $k$, $\n$, $\m$ and $\vave$ represent the average values of all components, respectively.
 
Figure~1 in \cite{Wang2020b} shows that if $\psi$ is not included, $k$ in Equation~\ref{eq:eb} is not a constant, but depends on the individual and total masses of BHs.
With the correction factor $\psi$, the value of $k$ becomes constant for most models with $\Mbh/M<0.4$.
The $N$-body models in \cite{Wang2020b} are low-mass clusters with only two mass components and no stellar evolution.
With more realistic models, we provide a similar analysis by treating BHs and non-BHs as two components.

\begin{figure}
  \includegraphics[width=0.95\columnwidth]{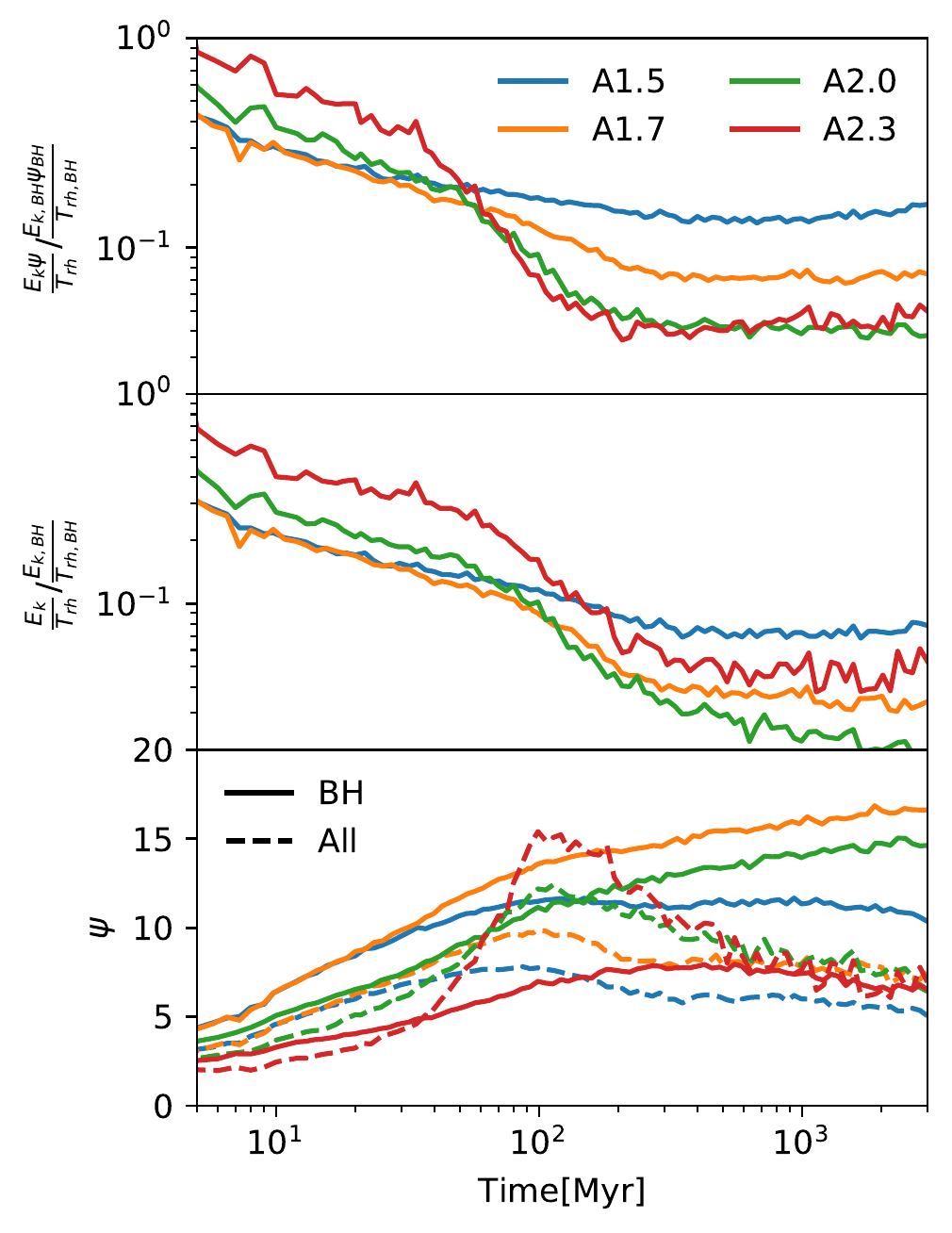}
  \caption{The evolution of energy flux rate (measured at $\rh$ and $\rhbh$) with (the upper panel) and without (the middle panel) the correction factor, $\psi$, respectively. $\psi$ measured at $\rh$ and $\rhbh$ are shown in the bottom panel.}
  \label{fig:et}
\end{figure}

Figure~\ref{fig:et} shows the evolution of energy flux rate (measured at $\rh$ and $\rhbh$) with and without the $\psi$ factor, and the evolution of $\psi$ of all objects and BHs, respectively.
With the correction of $\psi$ to $\trh$ (the upper panel), the A2.0 and A2.3 models show the same ratio of energy flux, while the A1.5 and A1.7 models have higher ratios.
Without correction (the middle panel), however, there is no common ratio among all four models.
The lower panel shows the evolution of $\psi$ measured at $\rh$ and $\rhbh$.
The value of $\psi[\rhbh]$ initially increases for all models but it increases more rapidly for the top-heavy models (the A1.5 and A1.7 models). 
Then, it slightly decreases after 1~Gyr for the A2.0 and A2.3 models. The value of $\psi[\rh]$ also increases in the beginning. For the A2.3 model, it significantly increases at around 100 Myr and then decreases to a similar value to $\psi[\rh]$.
For the other models, it also peaks at around 100~Myr but has a lower value compared to $\psi[\rhbh]$ later on.
The analysis here ignores the issue of the internal $\psi$ factors for the BH and non-BH components discussed in Section~\ref{sec:massloss}.
But the result is consistent with that reported by \cite{Wang2020b}.

\section{Discussion}
\label{sec:discussion}

In this work, we did not include the general relativity effect on BH orbits during the simulation, but considered it in the post-process analysis to detect the BBH mergers.  
Thus, the BBH mergers that occurred in a short time between two snapshots are missed in our analysis.
Such events can occur in chaotic interactions of triple or quadruple systems.
\cite{Kremer2019} and \cite{Samsing2020} show that GW capture during resonant encounters can contribute to the BH mergers in GCs. 
Our models also miss such events. 
Therefore, the number of BBH mergers in our analysis is the lower limit. 

Besides, we cannot detect hierarchical mergers, which can be the sources for massive BHs detected by LIGO/VIRGO \citep{Abbott2019,Abbott2020}.
For the same reason, in our simulations, we also did not find intermediate-mass black holes (IMBHs), which are discussed in \citep{PortegiesZwart2004,Giersz2015,Rizzuto2020}, and the possible tidal-encounter driven BBH mergers \citep{Fernandez2019}.

Without primordial binaries, these models probably underestimate the BBH merger rates, since primordial binaries can lead to stellar-evolution-driven formation of BBHs.  
The exchange of components via dynamical encounters between primordial binaries and BHs can also generate BBHs.  This channel is also missed in our current simulations.

By including the tidal field, the survival timescale of A1.5 and A1.7 can be much shorter. This can also affect the BBH merger rates. 
In the future work, we will develop new models by improving all these aspects.

Since we include PPSN in our simulations, the models with top-heavy IMFs have a large fraction of equal-mass BBHs ($40.5 M_\odot$).
This can be changed by adopting the metallicity that is different from $Z=0.001$ or different stellar-evolution models for massive stars.
Thus, our finding of the mass ratio distribution of BBHs cannot represent all kinds of conditions.
However, the trend, i.e., top-heavy IMFs tend to lead to a higher mass ratio of BBHs, would be general even if the stellar evolution models were changed.

Since the dynamical driven BBH mergers have a large initial eccentricity compared to the mergers via binary stellar evolution, the upcoming space-borne GW detectors (LISA and Tian-qin) may detect the high eccentricy BBH mergers that can help to distinguish the origins \citep[e.g.][]{Kremer2019,Liu2020}.
The top-heavy IMF results in more massive mergers of about $80 M_\odot$, which could be easier to detect by these GW detectors.

\section{Conclusions}
\label{sec:summary}

In this work, we carry out four star-by-star $N$-body simulations of GCs with different IMFs, and initially, with $M=5\times10^5 M_\odot$ and $\rh=2$~pc. We find that the formation rate of BBH mergers depends on the stellar evolution and dynamical process (core collapse of BH subsystems and BH heating) in a complicated way.
There is no monotonical correlation between the slope of IMF ($\alpha_3$) and the number of (potential) BBH mergers (Figure~\ref{fig:bbhat}, Table~\ref{tab:bbhevent}).
The stronger stellar-wind mass loss in the first 100~Myr leads to a faster expansion of GCs with more top-heavy IMFs.
As the escape velocity is lower in more top-heavy models, although the number of BHs is much higher, the BBH merger rate is lower.
However, the A1.5 model, which shows a deeper core collapse of the BH subsystem after expansion, can produce a burst of BBH merger candidates at the momentum of the core collapse (Figure~\ref{fig:bbhat},\ref{fig:rct}).

During the long-term evolution, it is difficult to form BBH mergers in GCs with more top-heavy IMFs because they expand faster.
This trend can be identified from the evolution of the minimum semi-major axis of hard BBHs shown in Figure~\ref{fig:bbhat}.
Therefore, with the same initial mass and size, GCs with more top-heavy IMFs less efficiently produce BBH mergers within the same time interval of evolution.

However, GCs with top-heavy IMFs may maintain the escape velocity high enough for a long term  to retain BHs inside clusters. As a result, the total number of BBH mergers can be large in the case of the top-heavy IMFs \citep{Weatherford2021}. In other words, although the efficiency is low, high-density GCs with top-heavy IMFs can take a longer time to produce BBH mergers.
In the case of GCs with top-light IMFs, however, the total amount of BBH mergers are limited to the total number of BHs, even though they are more efficient.

The comparison with the two-component models from \cite{Wang2020b} suggests that the mass loss rate of BHs does not simply depend on the mass segregation time ($\tms$), probably because it is difficult to define an accurate $\tms$ when a mass spectrum exists.
However, the general trend of energy balance (energy flux rate) is consistent with the result of \cite{Wang2020b}.
We also identify the two evolution trends of GCs.
BHs escape faster in GCs with the canonical \cite{Kroupa2001} IMF ($\alpha_3=-2.3$), while light stars are lost faster in the case of top-heavy IMFs (the A1.5 and A1.7 models).
The former can finally become dense GCs like those observed in the Milky-Way galaxy, while the latter become dark clusters with none or very few stars.
Since observations can only detect luminous GCs, the contributions of BBH mergers from dark clusters are ignored, but their contributions can be important, as is also discussed in \citep{Weatherford2021}.

\section*{Acknowledgments}
L.W. thanks the financial support from JSPS International Research Fellow (School of Science, The University of Tokyo). M.F. was supported by The University of Tokyo Excellent Young Researcher Program.
This work was supported by JSPS KAKENHI Grant Numbers 17H06360 and 19H01933 and MEXT as “Program for Promoting Researches on the Supercomputer Fugaku” (towards a unified view of the universe: from large scale structures to planets, revealing the formation history of the universe with large-scale simulations and astronomical big data).
Numerical computations were carried out on Cray XC50 at Center for Computational Astrophysics, National Astronomical Observatory of Japan.

\section*{Data availability}
The simulation data underlying this article are stored on Cray XC50.
The data were generated by the software \textsc{petar}, which is available in GitHub, at https://github.com/lwang-astro/PeTar.
The simulation data will be shared via  private communication with a reasonable request.

\appendix

\label{lastpage}

\end{document}